\documentclass[useAMS,usenatbib]{mn2e}
\usepackage{graphicx,amssymb}
\usepackage{mnras_macros}

\newcommand{\bm}[1]{\mbox{\boldmath $ #1 $}}
\newcommand{\dm}{{\rm d}}

\title[Satellite galaxies in simulations of galaxy formation]
{The properties of satellite galaxies in simulations of galaxy formation}
\author[T. Okamoto, C. S. Frenk, A. Jenkins, and T. Theuns]{Takashi
Okamoto$^{1, 2}$\thanks{E-mail: 
tokamoto@ccs.tsukuba.ac.jp}, Carlos S. Frenk$^{2}$, Adrian
Jenkins$^{2}$, Tom Theuns$^{2, 3}$\\ 
$^{1}$ Center for Computational Sciences, University of Tsukuba, 1-1-1 
 Tennodai, Tsukuba 305-8577 Ibaraki, Japan\\  
$^{2}$Institute for Computational Cosmology, Department of Physics,
Durham University, South Road, Durham, DH1 3LE\\
$^{3}$Department of Physics, University of Antwerp, Campus
Groenenborger, Groenenborgerlaan 171, B-2020 Antwerp, Belgium}
\begin{document}

\date{Accepted . Received ; in original form }

\pagerange{\pageref{firstpage}--\pageref{lastpage}} \pubyear{2009}

\maketitle

\label{firstpage}

\begin{abstract}
We investigate the properties of satellite galaxies in cosmological
$N$-body/SPH simulations of galaxy formation in Milky Way-sized
haloes. Because of their shallow potential wells, satellite galaxies
are very sensitive to heating processes which affect their gas
content. Their properties can therefore be used to constrain the
nature of feedback processes that regulate galaxy formation. In our
simulations, we assume that all the energy produced by supernovae is
used as kinetic energy to drive galactic winds. Several of our
simulations produce bright, disc-dominated galaxies. We find that wind
models in which the wind speed, $v_{\rm w}$, is proportional to the
local velocity dispersion of the dark matter, $\sigma$ (and thus the
wind mass-loading, $\eta_{\rm w} \propto \sigma^{-2}$), make star
formation in satellites sporadic, reproduce the observed satellite
luminosity function reasonably well (down to $M_v=-7$) and match the
luminosity-metallicity relation observed in the Local Group
satellites. By contrast, models that assume a constant wind speed
overproduce faint satellites and predict an incorrect
luminosity-metallicity relation. Our simulations therefore suggest
that the feedback processes that operate on the scale of satellite
galaxies should generate galactic outflows whose mass-loading varies
inversely with the depth of the potential.
\end{abstract}


\section{Introduction}

The `$\Lambda$-cold dark matter' ($\Lambda$CDM) model is now broadly
accepted as the standard paradigm in cosmology. This model completely
specifies the initial conditions for the formation of cosmic
structure, as well as the values of the cosmological parameters. Thus,
in principle, it is possible to calculate the predicted evolution of
objects of any kind, provided that the relevant astrophysical
processes are understood. The complexity of these processes makes
computer simulations the ideal tool to investigate the formation of
structure in the Universe.

There is a long history of attempts to simulate the formation of
spiral galaxies from $\Lambda$CDM initial conditions 
\citep[e.g.][]{aba03a, sgp03, gov04, rob04, oka05, gov07, ck09, sca09}. 
Early on an `angular momentum problem' was identified. In the
$\Lambda$CDM model, galaxies are generally assembled by mergers. As
fragments merge, their angular momentum, which is primarily invested
in their orbits, is transferred to the outer halo. Thus, if these
fragments already carry within them the majority of the baryons which
will make up the final galaxy, then these baryons will end up in a
slowly rotating central spheroid rather than in a disc
\citep{fwed85}. The likely solution to the problem was also identified
early on: the baryons must be kept from cooling into the fragments,
perhaps by some form of feedback, until the merging activity has
subsided after which they can cool smoothly to form a disc
\citep*{nb91, nw94, nfw95}.

The nature of the angular momentum problem and the key role of
feedback were explicitly demonstrated by \citet{oka05} who simulated
the formation of three galaxies -- all in the same dark matter halo --
assuming three different models of feedback powered by the energy
liberated in supernovae (SNe). The resulting three galaxies spanned the
whole range of morphological types, from a pure spheroid to a disc
galaxy with bulge-to-total mass ratio of 0.5. \citet{zof08} explored
these galaxies further and showed how, when the feedback is not strong
enough to prevent gas cooling early on, a spheroid forms, whereas
strong feedback can result in the formation of a disc by infall of gas
which approximately conserves its angular momentum.

The physical processes that establish the feedback loops required for
disc galaxies to form are poorly understood, but almost certainly
involve SNe energy and perhaps also energy extracted from a
central black hole \citep[e.g.][]{dsh05, oka05, sij07, bs09}. 
These processes occur on scales which are many orders of magnitude 
below those that can be followed directly in galaxy simulations 
from $\Lambda$CDM initial conditions. They must therefore be treated 
as `sub-grid' physics. A number of authors have attempted this kind 
of modelling using various recipes leading to a new generation of 
simulations which are producing increasingly realistic disc galaxies 
\citep{sh03, rob04, oka05, sti06, gov07, sca08, ck09}.

Although a full physical understanding of how galactic feedback
processes work seems a long way away, it is worth trying to identify
or constrain the `macroscopic' properties to which those processes
must give rise for a galaxy simulation to be successful. Since it
seems clear that much of the activity must take place early on, a
promising approach is to investigate how different feedback models
influence the properties of surviving fragments which appear today as
satellites of the main galaxy. This is the approach we take in this
paper: we focus on the luminosity function and metallicity of
satellites and examine how these properties are affected by different
models of feedback.

The properties of satellites are interesting in their own right. For
example, their luminosity function has often been claimed to pose a
major problem for the CDM model because the number of satellites seen
around the Milky Way is several orders of magnitude smaller than the
number of substructures that survive the collapse of galactic haloes in
$N$-body simulations \citep{kly99, moo99}. However, there are
various feedback processes that could explain why only a tiny fraction
of subhaloes manage to make stars. SN heating is clearly one of them 
\citep{ben02,ben03}; another is the photoionization of gas at early 
times, which suppresses gas cooling and raises the thermal pressure 
of the gas, preventing it from accreting onto small dark matter haloes 
\citep*{br92, efs92, tw96, qkf96, gne00, ogt08}.

These processes can be readily modelled using semi-analytic techniques
and such studies have shown that indeed they greatly reduce the number
of visible satellites (\citealt*{bkw00};
\citealt{ben02, som02, li09, mac09}). Some doubt has recently been cast on the
validity of these important conclusion because both the early studies
and the more recent, but similar ones by \cite{li09} and \cite{mac09},
were based on Gnedin's (2000) filtering mass description of
reionization which appears to overestimate significantly the
suppression of galaxy formation \citep{hoe06, ogt08}.
\citet{mac09}, however, have shown that even with the weaker suppression 
advocated by \citet{ogt08}, acceptable satellite luminosity functions
can still be obtained.

Hydrodynamic simulations are well suited for studying complex
processes such as reionization and SN feedback {\it ab initio}, with a
minimum of simplifying assumptions. {An earlier generation of galaxy
simulations showed that at least the brightest satellite galaxies are
produced in acceptable numbers \citep{gov07, lib07}. These
simulations, however, lacked the numerical resolution required to
study the bulk of the observed Milky Way satellites.}

{The simulations presented in this paper include a number of baryonic
processes known to be relevant to galaxy formation. In particular, we
model SN feedback as winds, a phenomenological approach which was
originally introduced by \citet{sh03}.  \citet{od06} modified this
treatment allowing the wind speed and mass-loading to vary with galaxy
properties, while \cite{onb08} introduced a model in which the energy
in the winds reflects the timed-release of SN energy. Unlike in
previous simulations, \citet{ds08} considered the case in which wind
particles are allowed to interact hydrodynamically with interstellar
medium (ISM) particles and found that due to hydrodynamical drag in
the ISM, both the {\it effective} wind speed and the {\it effective}
mass-loading (i.e. the wind speed and mass-loading when the wind leaves the
star-forming region) change significantly from the {\it input} wind
speed and mass-loading. This type of wind models are also used in
recent cosmological simulations \citep[e.g.][]{gimic, owls}.}

{Winds are now observed in many star-forming galaxies.  Earlier data
\citep{mar99, hec00} suggested that their properties were independent of
galactic properties. More recent observations \citep*{mar05, rvs05},
however, have reversed this view: galaxies with higher circular
velocity and higher star formation rates appear to drive faster winds.
In this paper, we consider two types of wind models, both of which
assume that all the energy released from SNe is converted into the
kinetic energy of the winds. In one case, the wind speed is assumed to
be constant, as in \citet{sh03} and as suggested by the earlier data
\citep{mar99}. In the other case, the wind speed is assumed to be
proportional to the local velocity dispersion, as suggested by the
more recent data \citep{mar05, rvs05}. Similar, but different wind
models have been used successfully to investigate the ejection of
intergalactic metals \citep{od06, od08} and the galaxy
mass-metallicity relation of massive galaxies \citep[$M_* \gtrsim 10^9
M_\odot$; ][]{fd08}.}

{This paper is organised as follows. In Section~2, we describe our 
simulations, providing brief descriptions of our modelling of baryonic
processes (Section~2.1), the energetic and chemical feedback due to
stellar evolution (Section~2.2) and the implementation of winds
(Section~2.3).  In Section~3 we present our results. We first focus on
the properties of the central galaxies (Section~3.1). We then address
how the luminosity functions and metallicities of the satellite
galaxies are affected by different feedback prescriptions
(Section~3.2). In Section~3.3 we perform convergence tests by using a
higher resolution simulation. Finally, in Section~4, we discuss the
results and summarise our main conclusions.}

\section[]{The Simulations}

We selected the three most massive objects from the sample of six
Aquarius haloes described in \cite{aquarius}.  The Aquarius haloes,
extracted from a large cosmological simulation, were chosen to be
relatively isolated and to have masses suitable to host a Milky
Way-like galaxy. With one exception, which is not in our sample, they
all have a relatively quiet merger history since $z=1$.  The three
most massive haloes are: Aq-A, Aq-C, and Aq-D.  Halo Aq-A has the
quietest merger history of the whole sample and has been studied in
the most detail to date. {All our simulations follow galaxy
formation within a periodic cube of side $100~h^{-1}$~Mpc in a
$\Lambda$CDM cosmology with parameters: $\Omega_{\rm m} =
0.25,~\Omega_\Lambda = 0.75,~\Omega_{\rm b} = 0.045,~\sigma_8 =
0.9,~n_{\rm s} = 1$, and $H_0 = 100~h~{\rm km}~{\rm s}^{-1}~{\rm
Mpc}^{-1} = 73~{\rm km}~{\rm s}^{-1}~{\rm Mpc}^{-1}$. These values are
the same as those adopted for the Millennium
Simulation~\citep{millennium} and the Aquarius project, and, within
the uncertainties, are consistent with the current set of constraints
from the WMAP 1-year
\citep{wmap1} and 5-year \citep{wmap5} data. 
}

Hydrodynamic simulations of galaxy formation in all six Aquarius
haloes have been presented by \citet{sca09}. These have similar
resolution to our simulations but use a different code and focus on
different issues. {They employed a slightly lower value of the
mean baryon density, $\Omega_{\rm b} = 0.04$, than us.}  In that work,
Aq-C, produced the most disc-dominated galaxy and had the smallest
mass fraction in subhaloes within $r_{50}$ (the radius inside which
the density is fifty times the critical density).  Aq-D has a similar
halo mass to Aq-A and Aq-C\footnote{Other Aquarius haloes have mass a
factor of $\sim 2$ smaller than these haloes}.  The concentration
parameter of this halo is the smallest of the three we have chosen
\footnote{{The values of the concentration parameter, $c$, for
haloes Aq-A, Aq-C, and Aq-D, obtained by fitting an NFW density
profile~\citep*{nfw97}, are $\sim 16$, $15$, and $9$, respectively
\citep{aquarius}.}}.

Since hydrodynamic simulations are much more computationally expensive
than $N$-body simulations with the same number of dark matter particles,
we are only able to simulate haloes at the lowest and next-to-lowest
resolution employed in \cite{aquarius} (corresponding to levels 5 and
4 in that paper). To test for numerical convergence, we carried out a
high resolution simulation of Aq-D, to which we refer as Aq-D-HR
(which corresponds to level 4).  
In Table~\ref{resolution}, we summarise the halo properties, the particle 
masses and the gravitational softenings used for the simulations.

\subsection{Baryonic processes}

The simulation code was developed from an early version of {\small
GADGET-3}. This is similar to the publicly available {\small GADGET-2}
\citep{spr05}, but it has a more efficient load-balancing algorithm.
We have modified {\small GADGET-3} to follow a number of additional
physical processes as follows.  

Both photo-heating by a spatially uniform, time-evolving ultraviolet
background and radiative cooling depend on gas metallicity, as
described in~\citet*[][]{wss09}, by assuming the gas to be optically
thin and in ionization equilibrium. We use the ultraviolet background
given by \citet{hm01} which is switched on at z = 9 (see also
\cite{ogt08}). {The cooling/heating routine described by 
\citet{wss09} computes the cooling/heating rates individually for eleven elements
whose contribution to the radiative cooling is significant (H, He, C,
N, O, Ne, Mg, Si, S, Ca, and Fe).  In the simulations, we track only
nine elements and we take S and Ca to be proportional to Si
\citep[see][]{wie09b}.}

{As in \citet{oka05}, we use the smoothed metallicity, rather than the
particle metallicity (which is normally used), to compute the
photo-heating and radiative cooling rates and to give the initial
amount of each individual chemical element to newly-born stars.
\citet{wie09b} found that the particle metallicity
underestimates the metallicity of relatively unenriched gas compared
to the smoothed value and thus significantly underpredicts the total
stellar mass formed in their cosmological simulations.
Using the particle metallicity is particularly problematic when a
small fraction of metal enriched particles are ejected as winds, as in
our simulations.}

{Star formation is modelled as in \citet{onb08}. That is, we treat
an SPH particle whose density exceeds a threshold density for star
formation ($n_{\rm H} > n_{\rm H,th} = 0.1~{\rm cc}^{-1}$) as a hybrid
particle that contains two distinct phases: hot ambient gas and cold
clouds, as in \citet{sh03}. In this model, the clouds have a fixed
mass spectrum and the star formation rate in a giant molecular cloud
is a function of the ambient pressure according to the two-phase model
of \citet{sg03}. \citet{booth07} introduced an alternative to the
hybrid particle approach in which the cold clouds are represented by
sticky particles (and hot, warm, and diffuse phases are represented by
SPH particles). While this model shares many physical processes with
our model, such as cloud formation by thermal instability and cooling
and evaporation by thermal conduction, it allows the cloud mass
spectrum to evolve with time through coagulation. }

Stellar evolution is modelled as in
\citet{onb08}, with two exceptions: (i) we employ the more realistic
Chabrier initial mass function \citep[IMF, ][]{chabrier03}, 
instead of the Salpeter IMF \citep{salpeter}; and (ii) we use 
metallicity-dependent stellar lifetimes and chemical yields 
\citep*{pcb98, marigo01}.  
The code does not assume the instantaneous recycling approximation.
The production of metals by SNe and AGB stars, stellar mass loss and
SN feedback all take place on the timescale dictated by stellar
evolution considerations. {It is however important to 
note that published nucleosynthesis yields and stellar lifetimes vary
significantly amongst authors.  A detailed discussion of this topic
may be found in \citet{wie09b}. } The stellar population synthesis
model P$\acute{\rm E}$GASE2
\citep{pegase} is used to calculate the luminosity of the simulated 
galaxies.

We now discuss the technical implementation of important ``subgrid''
processes included in our simulations.

\subsection{Supernova energy and mass loss}

In our simulations each star particle represents a simple stellar
population (SSP), specified by its IMF, age, and metallicity. All
stars more massive than $8 M_\odot$ are assumed to explode as
SNe~II. As the simulation proceeds, the supernova energy, mass loss,
and associated metals produced by a star particle are spread out over
neighbouring particles.  We use the lifetime and yields given by
\citet{pcb98}.  While it would be most natural to distribute
these quantities at each of the timesteps used to compute the
dynamics, this proves to be prohibitively expensive
computationally (particularly for SNe~Ia). 
We adopt instead a more economical stochastic scheme, introduced 
by \citet{onb08}, in which the distribution operation takes
place within a smaller number of discrete time intervals during the
lifetime of an individual star particle.

From the time a star particle is born until it reaches the age, $t_{8
M_\odot}$, corresponding to the lifetime of an $8 M_\odot$ star, 
({$\sim 40$~Myr for solar metallicity}), the distribution of mass,
metals and supernova energy to neighbouring particles occurs in a
series of discrete events set by a timestep, $\Delta t_{\rm II}$, of
magnitude a fiftieth of $t_{8 M_\odot}$. After an age $t_{8M_\odot}$
there are no more SNe~II, and a longer controlling timestep is
introduced: $\Delta t_{\rm Ia}$ = 100 Myrs. This longer step is chosen
specifically to follow SNe~Ia, but it is also sufficiently small to
allow the distribution of elements produced by AGB stars, as well as
stellar mass loss from the SSP as a whole.  In general, the dynamical
timestep, $\Delta t$, is much shorter than either $\Delta t_{\rm II}$
or $\Delta t_{\rm Ia}$. As a precaution, however, for star particles
with an age less than $t_{8 M_\odot}$, the dynamical time is
explicitly limited to a maximum of half of $\Delta t_{\rm II}$.

For a star particle with an age, $t < t_{8 M_\odot}$, we define the
conditional probability that the star particle undergoes a discrete
event resulting in the distribution of energy, mass, and metals during
a particular dynamical timestep, $\Delta t$, as follows:

\begin{eqnarray}
p_{\rm II} &=& \frac{\int_t^{t+\Delta t} r_{\rm II}(t') \dm t'}
  {\int_{t_0}^{t_0 + \Delta t_{\rm II}} r_{\rm II}(t') \dm t' - 
  \int_{t_0}^t r_{\rm II}(t') \dm t' } \nonumber \\
    &=& \frac{\int_t^{t+\Delta t} r_{\rm II}(t') \dm t'}
  {\int_t^{t_0 + \Delta t_{\rm II}} r_{\rm II}(t') \dm t'}, 
  \label{typeIIrate}
\end{eqnarray}
where $r_{\rm II}(t)$ is the SN~II rate for the SSP of age $t$ and
$t_0$ is the age when the previous SN~II timestep finished.  
{The SN~II rate, $r_{\rm II}(t)$, can be directly computed from the 
IMF, $\Phi(m) \equiv \dm N/\dm m$,  and the stellar lifetime, $t(m)$ as
\begin{equation}
r_{\rm II}(t) = - \Phi(m) \frac{\dm m(t)}{\dm t}
~{\rm for}~t(m_u) \le t \le t_{8 M_\odot}, 
\end{equation}
where the IMF is normalised as 
\begin{equation}
\int_{m_u}^{m_l} m \Phi(m) \dm m = 1, 
\end{equation}
and $m_u$ and $m_l$ are the upper and lower limits of the IMF
respectively. (We assume $m_u = 100~M_\odot$ and $m_l = 0.1~M_\odot$)
The total number of SNe~II produced by an SSP of $1~M_{\odot}$ is
$1.18 \times 10^{-2}$.
} 
We generate a uniform random number between zero and one at every $\Delta
t$. When $p_{\rm II}$ exceeds this number, the energy, mass, and
metals expelled by the SSP during the interval $t_0$ to $t_0 + \Delta
t_{\rm II}$ are distributed over the neighbouring gas particles and
the probability is set to zero until the end of this SN~II
timestep\footnote{For example, the energy released by an SSP of $1 M_\odot$
during $\Delta t_{\rm II}$ is $E_{\rm SN} \int_{t_0}^{t_0 + \Delta t_{\rm
II}} r_{\rm II}(t') \dm t'$, where $E_{\rm SN}$ is the energy released
by each supernova which, throughout this paper, we take to be $E_{\rm
SN} = 10^{51} \ {\rm erg}$}.

The energy, metals and mass produced by a particle are distributed
amongst its 40 nearest neighbouring gas particles at each of these
discrete events. It is important that each quantity that is being
redistributed be accurately conserved. To ensure that this happens, the
fraction of each quantity received by the $i$-th gas particle
neighbouring the star is given by $m_i/\sum_j m_j$, where the
subscript $j$ runs over all 40 neighbours.  We find that the results
are relatively insensitive to the exact number of neighbours, with
32--128 particles producing similar behaviour.

The same procedure applies after an age $t_{8 M_\odot}$, except that
the timestep is now $\Delta t_{\rm Ia}$. {We calculate the SN~Ia
rate using the scheme of \citet{gr98} with the updated parameters of
\citet{pcb98}. We take chemical yields and stellar mass loss from
intermediate mass stars from \cite{marigo01}. Further details may be
found in \citet{oka05}, \cite{naga05} and references therein.}

In summary, our method for treating feedback and mass loss allows us
to perform the expensive neighbour search only once per star
particle per SN timestep and results in the correct distribution
of SNe in time, even with large SN timesteps, when averaged over many
particles.  Note that the method guarantees a single distribution event
per star particle per SN timestep.

\subsection{Galactic winds}

In our subgrid model of feedback, SN explosions give rise to a wind
by imparting kinetic energy to nearby gas particles. We assume that
all of the energy released from SNe is potentially available to power
the kinetic energy of the wind. 
Hence in terms of feedback efficiency, there is no difference among 
our models. 

In order to prevent gas particles outside star-forming regions from
being launched as winds\footnote{This can occur, for example, if a
SN~Ia is produced by a star in a low density gas environment.}, we
assume that only star-forming gas particles ($n_{\rm H} > n_{\rm H,
th} = 0.1$ cm$^{-3}$, where $n_{\rm H, th}$ is the threshold density
above which star formation can occur) are eligible to become wind
particles.  If low density gas particles ($n_{\rm H} < n_{\rm H, th}$)
receive feedback energy, we simply add it to their internal energy.

During any given timestep, an eligible gas particle may receive
supernova energy, $\Delta Q$, from one or more neighbouring star
particles. If this happens, then the particle is selected to become a
wind particle during that timestep with a probability:
\begin{equation}
p_{\rm w} = \frac{\Delta Q}{\frac{1}{2} m_{\rm SPH} v_{\rm w}^2}, 
\label{p_w}
\end{equation}
where $m_{\rm SPH}$ is the mass of the gas particle and $v_{\rm w}$ is
the initial wind speed. The value of $v_{\rm w}$ is discussed later in this
subsection. Having decided whether a particle is to become a wind
particle or not, we set its value, $\Delta Q = 0$.

{This procedure ensures that the kinetic energy is injected {\it
locally} and that the generation of winds is consistent with the timed
release of SN energy and metals. This is an important difference from
the widely used wind model of \citet{sh03} in which the wind particles
are selected stochastically from all the star-forming gas (i.e. with
$n_{\rm H} > n_{\rm H,th}$) in the simulation and are therefore not
local to young star particles\footnote{Note however that in
\citet{sh03} the gas particle metallicity is updated according to its
star formation rate; hence the non-local winds are consistent with
the chemical enrichment scheme.}.
A different implementation of locally generated winds can be found
in \citet{ds08}. }

It is possible for $p_{\rm w}$ to exceed unity. We deal with this
situation using an iterative procedure.  Firstly, the particle is
added to the wind and we record both its label and the value of its
`excess' energy, $\Delta Q - \frac{1}{2} m_{\rm SPH} v_{\rm w}^2$.
Having considered all eligible gas particles to decide which become
wind particles, we then revisit those with excess energy and
redistribute this {energy} to neighbouring non-wind gas particles.
We then consider those particles whose density makes them eligible to
be selected as a wind particle, recompute $p_{\rm w}$ including the
reassigned excess energy from $\Delta Q$ in Eqn.~\ref{p_w}, and decide
whether they are converted into wind particles or not in the usual
way.  If there are still values of $p_{\rm w}$ which exceed unity, a
further iteration is begun, and these iterations continue until there
is no more excess energy to redistribute.  In practise no more than
two iterations are ever required.

When a particle is selected as a wind particle, its velocity, ${\bm
v}$, is incremented as:
\begin{equation}
{\bm v'} = {\bm v} + v_{\rm w} \hat{\bm n}. 
\label{direction-eq}
\end{equation}
The unit vector $\hat{\bm n}$ is chosen at random to be either
parallel or anti-parallel to the vector $({\bm v} - \bar{\bm v})
\times {\bm a}_{\rm grav}$, where $\bar{\bm v}$ ideally would be
chosen to be the velocity of the centre of mass of the galaxy. Since
it is difficult to determine this velocity as the simulation proceeds,
we instead select the average velocity of the neighbouring 40 dark
matter particles as a proxy.  The gravitational acceleration, ${\bm
a}_{\rm grav}$, on the other hand, includes contributions from matter
everywhere in the simulation. In practise, for the central regions of
haloes and subhaloes, both the magnitude and direction of ${\bm
a}_{\rm grav}$ are largely determined by the local mass distribution.
Adding a velocity according to Eqn.~\ref{direction-eq} leads to wind
particles being ejected preferentially along the rotation axis of a
spinning object, thus generating an `axial wind' \citep{sh03}. 
{Note that while the momentum is not strictly balanced 
when a single particle is placed in the wind, the momentum and angular 
momentum are statistically conserved because of the random orientation of the
momentum kick along direction $\bm{n}$. }

Since the physical mechanisms that drive the outflows in galaxies are
not well understood and the mass in winds is hardly constrained by
observations, we adopt two phenomenological models for the properties
of the winds (both of which assume that all of the supernova energy is
available to power the outflows).  The first model (which we label
`vw' for `variable wind') assumes that the initial wind speed, $v_{\rm
w}$, is proportional to the one-dimensional velocity dispersion,
$\sigma$, determined from the 40 neighbouring dark matter particles of
the gas particle in question. That is, in the 'vw' model, $v_{\rm w}
\propto \sigma$, as suggested by recent data \citep{mar05}. Here, we are
taking the local velocity dispersion as a proxy for the circular
velocity.  We have checked that the local velocity dispersion measured
at the position of a star-forming gas particle is strongly correlated
with the maximum of the circular velocity profile of its host
(sub)halo, $V_{\rm max}$, and that the relation between them 
($V_{\rm max} \simeq 1.45 \sigma$) does not evolve with redshift.  

{In order to characterise the `vw' models, we discuss wind
properties by assuming the instantaneous recycling approximation and
considering only SNe~II as the energy source for the winds. While we
do not make these assumptions in the simulations, considering this
simplest of cases is instructive. With these assumptions, the energy
ejection rate, $\dot{E}_{\rm SN}$, is proportional to the star
formation rate, $\dot{M}_*$. Therefore, the mass flux in the winds in
the `vw' models, $\dot{M}_{\rm w}$, can be written as:
\begin{equation}
\dot{M}_{\rm w} = 2 \frac{\dot{E}_{\rm SN}}{v_{\rm w}^2} \propto \frac{\dot{M}_*}{\sigma^2}, 
\end{equation}
because the wind speed, $v_{\rm w}$, is proportional to the local dark
matter velocity dispersion, $\sigma$. From these considerations, the
wind mass-loading, $\eta_{\rm w}$, can be expressed as:
\begin{equation}
\eta_{\rm w} \equiv \frac{\dot{M}_{\rm w}}{\dot{M}_*} =
\left(\frac{\sigma}{\sigma_0}\right)^{-2},  
\label{vw}
\end{equation}
where $\sigma_0$ is a parameter defined by the IMF and the feedback
efficiency. We give the values of $\sigma_0$ in our models in
Table~\ref{params}. As mentioned earlier, we adopt the Chabrier IMF
and make use of all the SN energy to power the kinetic energy of the
winds.}

For the second wind model (which we label `cw' for `constant wind'),
we assume a constant initial wind speed, which implies a constant
mass-loading. This type of wind model has been widely used in
hydrodynamic simulations of galaxy formation (e.g. \citealt{sh03};
\citealt*{nsh04}; \citealt{onb08}).  

{In both the `vw' and `cw' models, we assume that newly launched
wind particles are {\it decoupled} from hydrodynamical interactions
for a brief period of time in order to enable the winds to emerge from
a location close to the surface of their star-forming region with the
{\it specified} wind velocity and mass-loading. This wind decoupling
has been widely used in simulations since it was introduced by
\citet{sh03}.}

{
\cite{ds08} compared {\it coupled} (i.e. not decoupled) and 
{\it decoupled} winds in detail and found that coupled winds blow
bubbles and drive turbulence or create channels in the gas disc, while
decoupled winds remove fuel for star formation without disturbing the
gas disc. More importantly, for a given set of (input) wind speed and
mass-loading, coupled winds are more efficient at suppressing star
formation in low mass galaxies because they drag gas along;
consequently the mass-loading when they leave the star-forming region
(i.e. the effective mass-loading) is much larger than the input one.
On the other hand, coupled winds are less efficient in high mass
galaxies because they suffer large energy losses due to drag in the
high-pressure ISM and the decelerated winds cannot escape from the
galaxy (or often even from the star-forming region)~\citep{owls}.}



Since we wish to investigate the effects of two different feedback
models on the population of satellite galaxies, in which wind speed 
and mass-loading depend differently on the galaxy properties, 
we employ {\it decoupled} winds in this paper in order to specify 
the {\it effective} wind velocity and mass-loading as inputs. 
Full hydrodynamical interactions are enabled once a wind particle 
leaves the star-forming region ($n_{\rm H} < 0.1 n_{\rm H,
th}$) or after the time, $10 \ {\rm kpc}/v_{\rm w}$, has elapsed,
whichever occurs earlier\footnote{In our simulations, the first 
condition ($n_{\rm H} < 0.1 n_{\rm H, th}$), is usually satisfied
first.}.

\subsection{Models} 

\begin{table*}
\caption{Numerical parameters relating to our four sets of initial
 conditions. The virial mass, $M_{\rm vir}$, is defined as the mass
 within a sphere whose mean density is equal to the virial density at
 redshift zero computed from the spherical collapse model
 \citep{ecf96}. The gravitational softening lengths, $\epsilon$, are
 kept fixed in comoving coordinates for $z > 3$; thereafter they are
 frozen in physical units and to values presented in the table. }
\label{resolution}
\begin{tabular}{@{}lcccc}
\hline
Halo 
  & $M_{\rm vir}$ ($h^{-1} M_{\odot}$)  
  & $m_{\rm DM}$ ($h^{-1} M_{\odot}$)  
  & $m_{\rm SPH}$ ($h^{-1} M_{\odot}$)  
  & $\epsilon$ ($h^{-1}$ kpc)  \\ 
\hline
Aq-A & $1.4 \times 10^{12}$ & $1.9 \times 10^6$ & $4.1 \times 10^5$ & 0.425 \\
Aq-C & $1.2 \times 10^{12}$ & $1.5 \times 10^6$ & $3.4 \times 10^5$ & 0.425 \\
Aq-D & $1.3 \times 10^{12}$ & $1.9 \times 10^6$ & $4.1 \times 10^5$ & 0.425 \\
Aq-D-HR & $1.3 \times 10^{12}$ & $1.6 \times 10^5$ & $3.5 \times 10^4$ & 0.175 \\
\hline
\end{tabular}
\end{table*}
For the variable wind models, we show results for $v_{\rm w}
= 5 \sigma$ and $v_{\rm w} = 4 \sigma$; we refer to these as
`vw5$\sigma$' and `vw4$\sigma$', respectively.  Although we also
calculated a model with $v_{\rm w} = 3 \sigma$, we do not show the 
results here because this wind speed is insufficient to allow gas particles
to escape from their host halo and thus star formation is hardly suppressed
compared to the no feedback case \citep{onb08}.

For the constant wind models, we show results for $v_{\rm w} = 700$
and 600 km s$^{-1}$; we label them `cw700' and `cw600', respectively.
We employ a slightly faster wind speed than was used in
\citet{onb08} ($v_{\rm w} = 500$ km s$^{-1}$) because we find that this
wind speed is too low to suppress the early starbursts that would turn
our galaxies into bulge-dominated objects \citep{oka05}. We believe
that this is due to the fact that the baryon fraction in our current
models is 35\% higher than in the cosmological model adopted by
\citet{onb08}.  The values of the wind mass-loading factor, $\eta_{\rm w}$,
for these models are given in Table~\ref{params}.

For the models introduced so far, we employ a multiphase model for the
star-forming gas, as described in \citet{onb08}, in which a minimum
pressure for star-forming gas, $P_{\rm min} \propto \rho^{1.4}$, is
imposed in order to stabilise gas discs against gravitational
instability.  {Such a stiff equation of state for the star-forming
gas was first introduced by \citet{sh03} in an attempt to allow for
the fact that the interstellar medium (ISM) is supported by a hot gas
phase which is unresolved in the simulations.  On the other hand,
\citet{wn01, wn07} suggested that the multi-phase ISM is supported by
turbulent motions induced by the self-gravity of the gas and galactic
rotation. In any case, \citet{rob04} argued that a stiff equation of
state for the star-forming gas is required to form a disc. Our
adoption of such an equation of state is intended to allow for these
unresolved and ill-understood processes.  
}

{
Artificial fragmentation of a disk may occur when simulations do not
properly resolve the local Jeans length~\citep{truelove97, bb97}.  In
order to prevent this from happening, several authors have adopted a
similar strategy to us by imposing a minimum pressure for high-density
gas \citep{rk08, amanogawa1, cev09, gimic, owls}. }

{
\citet{owls} found that the global star formation rate is almost 
completely independent of both the equation of state of the ISM and
the star formation law because the star formation is highly self-regulated
by the winds. } We therefore perform simulations without this
multiphase model for comparison.  In this case, the star-forming gas
can cool down to the equilibrium temperature set by the balance
between photoheating and radiative cooling. Note that metal enriched
gas can cool far below $10^4$K because we include metal-line cooling
\citep{wss09}. We use this single-phase gas in a `vw5$\sigma$' wind
simulation and label it `vw5$\sigma_{\rm single}$'.

\begin{figure}
\begin{center}
\includegraphics[width=8.5cm]{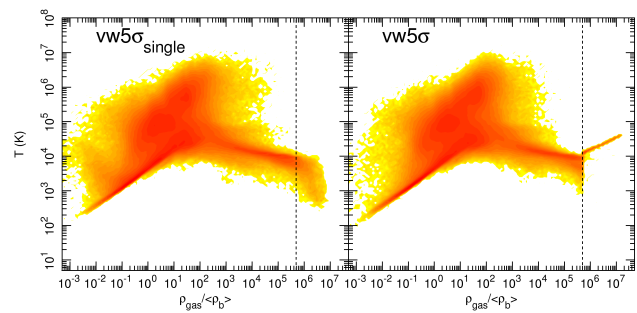}
\end{center}
\caption{ Distribution of gas particles in the density-temperature phase 
  diagram for the Aq-A halo at $z = 0$.  The left and right panels show the 
    `vw5$\sigma_{\rm single}$' and `vw5$\sigma$' models, respectively. 
    The density above which star formation is enabled is indicated by the 
    vertical dotted line. The ISM is well enriched 
    by $z = 0$. As a result there is low-temperature ($< 10^4$ K)
    star-forming gas in  
    the `vw5$\sigma_{\rm single}$' model which has been cooled by metal-line
    cooling.  We show the effective temperature for the multiphase 
    star-forming gas in the `vw5$\sigma$' model. Most of the star-forming gas 
    lies on the line specified by $P_{\rm min} \propto \rho^{1.4}$.  
}
\label{phase}
\end{figure}

In Fig.~\ref{phase} we compare the gas distribution in the
density-temperature phase diagrams for the `vw5$\sigma_{\rm single}$'
and `vw5$\sigma$' simulations in the Aq-A halo at $z = 0$.  Whereas
the star-forming gas in the `vw5$\sigma_{\rm single}$' model cools
down to $\sim 100$K because of metal-line cooling and an increasing
cooling rate with density, the gas in the `vw5$\sigma$' model lies on
the imposed minimum temperature. The equation of state of the
star-forming gas in the `vw5$\sigma$' model is hence much stiffer than
that in the `vw5$\sigma_{\rm single}$' model.  For both ISM models,
the star formation efficiency is normalised to reproduce the observed
relation between the surface star formation rate density and the
surface gas density \citep{ken98}.
\begin{table}
\caption{Wind models and their properties. $v_{\rm w}$ and $\eta_{\rm w}$
denote the initial 
wind speed and the wind mass-loading (see text).}
\label{params}
\begin{tabular}{@{}lccc}
\hline
Model
 & $v_{\rm w}$ (km s$^{-1})$
 & $\eta_{\rm w}$
 & Multiphase \\
\hline
vw5$\sigma_{\rm single}$ 
  & $5 \sigma$ & $\left(\frac{\sigma}{217 {\rm km} \ {\rm s}^{-1}}\right)^{-2}$ & No \\
vw5$\sigma$ 
  & $5 \sigma$ & $\left(\frac{\sigma}{217 {\rm km} \ {\rm s}^{-1}}\right)^{-2}$ & Yes \\
vw4$\sigma$ 
  & $4 \sigma$ & $\left(\frac{\sigma}{271 {\rm km} \ {\rm s}^{-1}}\right)^{-2}$ & Yes \\
cw700 
  & 700 & 2.4 & Yes \\
cw600 
  & 600 & 3.3 & Yes \\
\hline
\end{tabular}
\end{table}
%

\section{Results} 

We begin this section by presenting an overview of the properties of
the central galaxies. {While our main focus is on the satellite
galaxy population, the bulk properties of the central galaxies,
including their morphology, are a useful aid in understanding the
nature of our wind models which, in turn, are the key to understanding
the properties of the satellites~\citep{md02, oka05, zof08}. } A more
detailed analysis of the central galaxies will be presented in a
forthcoming paper.

\subsection{The central galaxies} 

\begin{figure}
\begin{center}
\includegraphics[width=8.5cm]{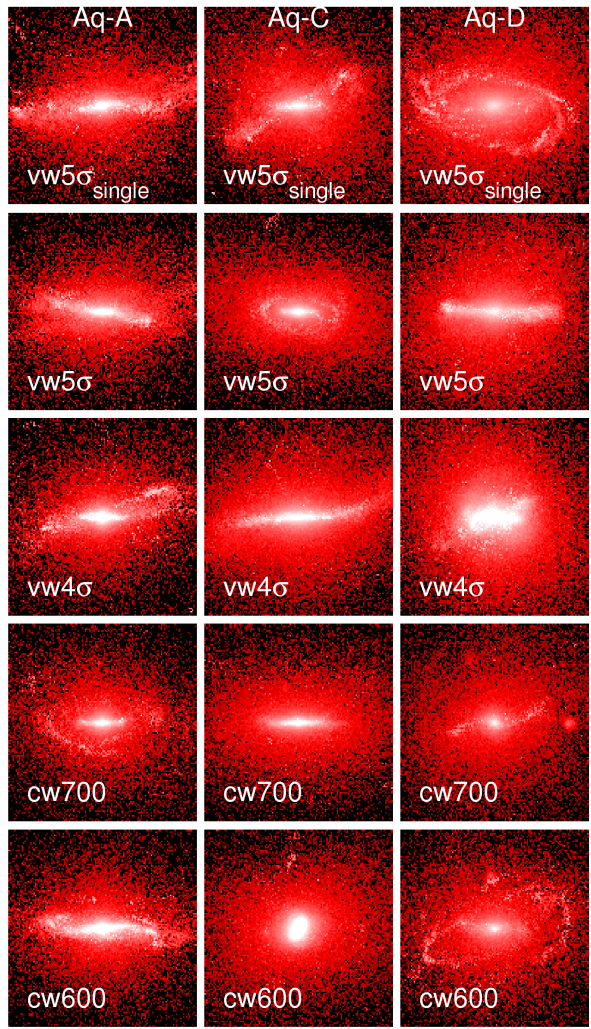}
\end{center}
\caption{Edge-on views of the central galaxies at $z = 0$. 
 The brightness is coded by the $B$-band surface brightness within a
 $50 h^{-1}$ kpc box centred on each galaxy.  The viewing angles are
 determined using the angular momentum of stars within 5\% of the
 virial radii of the host haloes. The Aq-A, Aq-C, and Aq-D haloes are
 shown from left to right, and models, `vw5$\sigma_{\rm single}$',
 `vw5$\sigma$', `vw4$\sigma$', `cw700', and `cw600' are shown from top
 to bottom.  }
\label{snapshots}
\end{figure}
A visual impression of the structure of our simulated galaxies may be
gained from Fig.~\ref{snapshots} which shows edge-on views of the
galaxies at $z=0$. To make these images, we define a `$z$' axis for each
central galaxy as the direction of the angular momentum of all stars
within $0.05 R_{\rm vir}$ of the centre at $z=0$. We take $R_{\rm
vir}$ to be the radius of a sphere that has the virial overdensity
given by the spherical collapse model \citep{ecf96}.  We define
two mutually perpendicular axes, `$x$' and `$y$', at random for each
simulation, perpendicular to the $z$-axis.  Fig.~\ref{snapshots} shows
that striking differences in the morphologies of the galaxies arise in
each of the haloes Aq-A, Aq-C and Aq-D, when the wind prescription is
varied. This supports the claim of \cite{oka05} that the morphology of
simulated galaxies is very sensitive to the details of the physics
implemented in the simulation. For example, for halo Aq-C a mere
change of $\sim 15\%$ in the wind speed from 700 km s$^{-1}$ to 600 km
s$^{-1}$ results in a change from a disc-dominated galaxy to a
spheroidal galaxy. Nonetheless, it is encouraging that some models do
successfully form disc-dominated galaxies.

\begin{figure}
\begin{center}
\includegraphics[width=8.5cm]{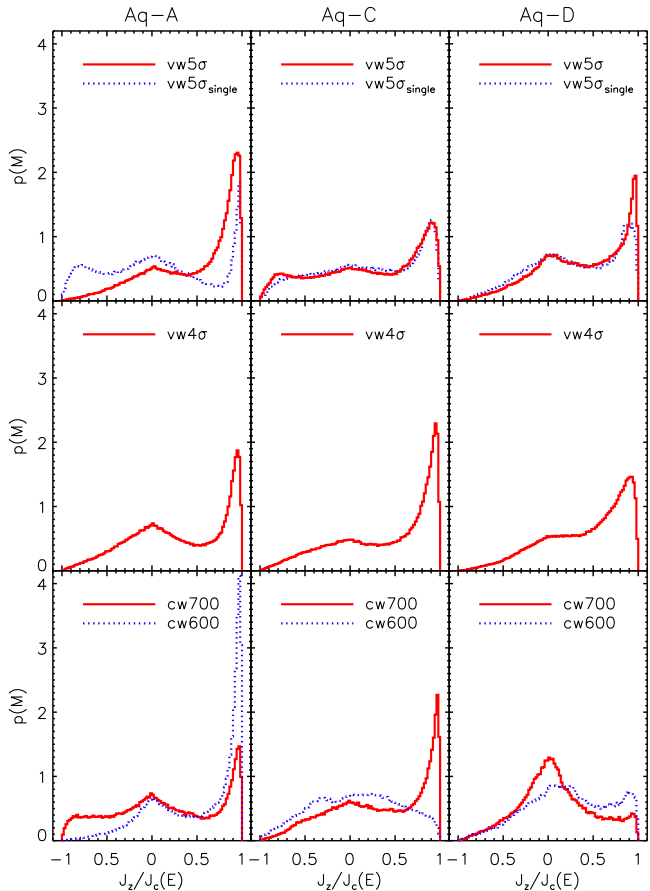}
\end{center}
\caption{The mass-weighted normalised distribution of orbital
 circularity, $J_z/J_{\rm c}(E)$, of all stars at $z=0$ lying within
 $R_{\rm vir}$ of the centre, excluding stars identified as members of
 satellite galaxies.  The Aq-A, Aq-C, and Aq-D haloes are shown from
 left to right and the `vw5$\sigma$', `vw4$\sigma$', and `cw' models
 are shown from top to bottom.  A peak at $J_z/j_{\rm c}(E) \simeq 1$
 indicates the existence of a disc component.  The `vw5$\sigma_{\rm
 single}$' and `cw700' models of the Aq-A halo and the `vw5$\sigma$'
 model of the Aq-C halo exhibit a retrograde disc component as well as
 a prograde one.  }
\label{morphology}
\end{figure}

The morphology of a galaxy is largely dictated by the angular momentum
distribution of the stars.  To quantify this, we analyse the
distribution of the orbital circularity, defined below, of stars in
the central galaxies. We exclude stars belonging to the satellites
(which we define precisely in the next subsection).  For each star
belonging to a central galaxy we compute, $J_z$, the component of
specific angular momentum parallel to the $z$-axis defined earlier.
We then compute the specific angular momentum, $J_{\rm c}(E)$, of a
prograde circular orbit with the same binding energy as the particle.
{To evaluate $J_{\rm c}(E)$, we first compute the circular velocity
profile of the host halo, $v_{\rm c}(r)$. The specific binding energy
of a particle on a circular orbit at radius $r$ is then given by
\begin{equation}
E(r) = - \frac{G M(< r)}{r} + \frac{1}{2} v_{\rm c}(r)^2.  
\end{equation}
Similarly, the specific angular momentum of a particle at $r$ is:
\begin{equation}
J_{\rm c}(r) = r v_{\rm c}(r). 
\end{equation}
We then calculate the specific energy of the $i$-th star particle assuming a 
spherically symmetric halo:
\begin{equation}
E_i = - \frac{G M(< r_i)}{r_i} +  \frac{1}{2} {\bm v}_i^2. 
\end{equation}
We identify the radius, $r$, of the circular orbit of specific binding
energy, $E(r) = E_i$ using a Newton-Raphson method. This then gives
the value of $J_{\rm c}(E_i)$.
}
The ratio $J_z/J_{\rm c}(E)$ defines the orbital circularity 
\citep{aba03b, oka05}.

Fig.~\ref{morphology} shows the normalised distribution of the orbital
circularity of stars in the central galaxy within $R_{\rm vir}$. A
cold disc component has $J_z/J_{\rm c}(E)\simeq 1$ and such a
component is evident in most of the galaxies except in the `cw600'
model of the Aq-C halo. {The `cw600' model produces the most
disc-dominated galaxy in the Aq-A halo, but the least disc-dominated
galaxy in the Aq-C halo. For a fixed wind model, halo-to-halo
differences in the final galaxy morphology reflect differences in the
formation and merging histories of the haloes (even though most of
them have similar quiet histories at early times). Galaxy morphology
is particularly sensitive to formation history in the `cw' models
because of the introduction of a critical velocity scale for the
winds\footnote{{
\bf This is not true for coupled winds \citep[see][]{ds08}.  
}}. In the `vw' case, the dependency on halo formation history is much
weaker. }

In Aq-A and Aq-D, the most pronounced discs among the `vw' models
form in the `vw5$\sigma$' model, but in Aq-C, it is the `vw4$\sigma$'
which produces the largest disc. 
The `vw5$\sigma_{\rm single}$' galaxies are less disc-dominated than 
their `vw5$\sigma$' counterparts. 
That confirms an earlier claim that the stiffer equation
of state of the star-forming gas helps disc formation
\citep{rob04}. The effect is, however, not always evident (see Aq-C). 
Interestingly, the `vw5$\sigma_{\rm single}$' and `cw700'
models of the Aq-A halo and the `vw5$\sigma$' model of the Aq-C halo
exhibit counter-rotating or retrograde discs in addition to the
prograde discs.  The origin of these counter-rotating discs will be
discussed in a forthcoming paper.

Our `vw' galaxies are more disc-dominated than those simulated by 
\citet{sca09} in the same Aquarius haloes, using very similar initial
conditions but a different galaxy formation code. Since galaxy
morphology is very sensitive to the details of feedback, this
discrepancy is not surprising and probably reflects the fact that the
strong feedback in the `vw' models is able to suppress star formation
in small haloes more efficiently than that in their simulations.

\begin{figure}
\begin{center}
\includegraphics[width=8.5cm]{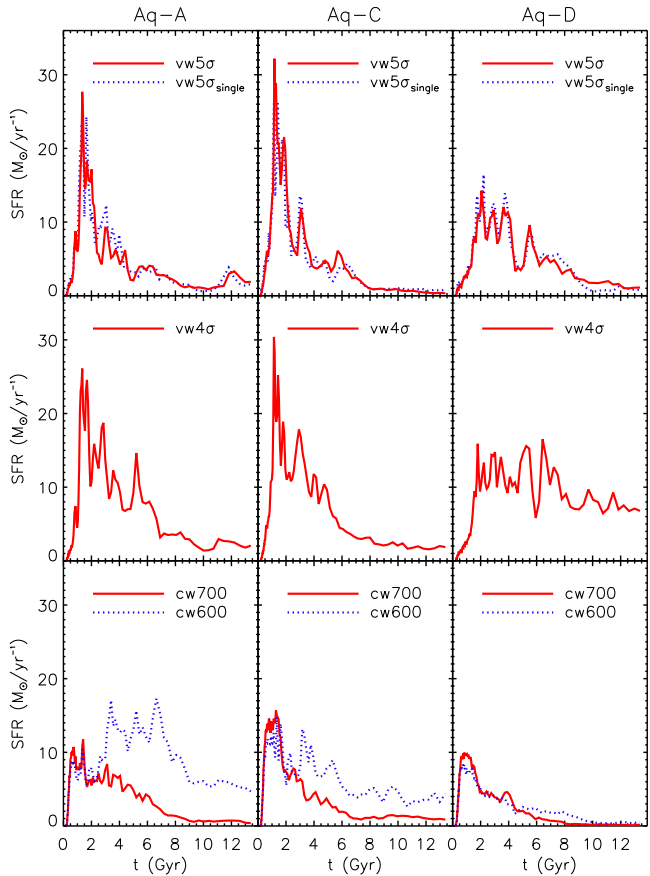}
\end{center}
\caption{The star formation histories of the central galaxies. Only
  stars within $0.1 R_{\rm vir}$ of the central galaxy which are not
  members of a satellite galaxies are shown.  The panels and lines are
  the same as in Fig.~\ref{morphology}.  The `vw$5\sigma_{\rm
  single}$' and `vw$5\sigma$' models show almost identical star
  formation histories, implying that the modelling of the ISM has
  little effect on the star formation histories and that it is the
  wind model which is the most important ingredient in determining the
  morphology of the galaxy.}
\label{sfh}
\end{figure}
As emphasised by \cite{oka05}, the star formation history of a galaxy
is correlated with its morphology. Star formation histories for the
central galaxies in our simulations are shown in Fig.~\ref{sfh}.  The
`vw5$\sigma$' and `vw5$\sigma_{\rm single}$' models are very similar,
indicating that the star formation history is even less sensitive to
the adopted ISM model than the morphology of the galaxy. {
This is consistent with the conclusions of \citet{owls} who find that the
global star formation is highly self-regulated by feedback and is
almost independent of the choices of equation of state for the ISM and
star formation law. }

The `vw5$\sigma$' and `vw4$\sigma$' models are also similar except that
the `vw4$\sigma$' model for the Aq-D halo has significantly higher
star formation rate at low redshift.  It is interesting that the
`cw600' model of the Aq-C halo has a relatively high star formation
rate at low redshift, a feature which was claimed to be a sign of disc
formation \citep{oka05}, although in this case this galaxy is
dominated by a spheroidal component (see Fig.~\ref{snapshots} and
\ref{morphology}).  Note that our sample of Aquarius haloes all have
quiet merger histories at low redshift. As recently stressed by
\cite{sca09}, the origin of galaxy morphology is more complicated
than we generally thought and the star formation history is only a very
crude indicator of the final morphology. 

\subsection{The satellite galaxies} 

\begin{figure}
\begin{center}
\includegraphics[width=8.5cm]{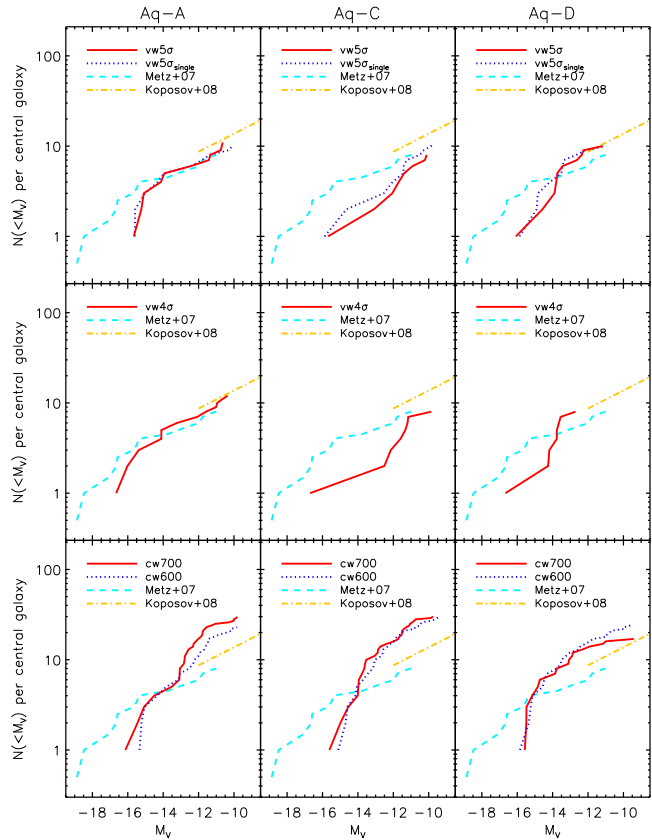}
\end{center}
\caption{Satellite luminosity
  functions. Only satellites within 280 kpc from the central galaxies
  are counted. The panels are the same as in
  Fig.~\ref{morphology}. The solid lines show the results from the
  simulations. The dashed line shows the luminosity function for the
  bright ($M_V < -11$) Milky Way and M31 satellites, similarly taking
  only those within 280 kpc from either galaxy, constructed from the
  data compiled by \citet{mkj07}.  The dot-dashed line is a power-law
  fit to the luminosity function of \citet{kop08} which includes newly
  discovered SDSS ultra-faint satellites, again
  restricted to include only satellites within 280 kpc of 
  the Milky Way.  }
\label{lf}
\end{figure}
We now turn to an investigation of how different feedback models
affect the properties of satellites. We identify subhaloes using the
{\small SUBFIND} algorithm \citep{spr01} and restrict attention to subhaloes of
at least 32~particles. In counting the number of particles in a
subhalo, we include not only dark matter but also baryonic
particles. Therefore it is, in principle, possible to identify purely
baryonic gravitationally bound objects. We define satellite galaxies
as subhaloes that contain at least 10~star particles. In Sec.~3.3 we
check the robustness of this definition and carry out a convergence
test.

\subsubsection{The satellite luminosity function} \label{section-lf}

In Fig.~\ref{lf}, we plot the satellite luminosity functions in our
simulations and compare them with observations of the Local Group. The
observed luminosity function of bright ($M_V < -11$) satellites lying
within 280~kpc of either the Milky Way or M31 is constructed from the
data compiled by \citet*{mkj07}.  The data of \citealt{kop08}, which
includes the incompleteness-corrected SDSS ultrafaint satellites
within 280 kpc from the centre of the Milky Way, are represented by a
power-law. The host galaxy haloes, defined by the friends-of-friends
algorithm \citep{defw85}, extend beyond 280 kpc, but, for consistency
with the observations, in constructing the luminosity functions we
consider only the satellites within 280 kpc of the centre. For other
analyses below we will make use of larger samples of satellites.

For a given wind model, the satellite luminosity function varies
significantly from halo to halo, reflecting the variation in the
number of massive subhaloes \citep{aquarius, ish09}.  This alerts us
to the danger of assuming that the Milky Way is a typical galaxy.  The
`vw5$\sigma_{\rm single}$' and `vw5$\sigma$' models again yield almost
identical results. Interestingly, the `cw700' and `cw600' models also
produce results almost indistinguishable from each other. The
`vw5$\sigma$ and `vw4$\sigma$' models are similar except that the
brightest satellite in a `vw4$\sigma$' model is slightly brighter than
its counterpart in the corresponding `vw5$\sigma$' model reflecting
the lower wind speed in the former model.

Our results are consistent with the conclusion of the semi-analytic
study by \citet{ben02} that it is difficult to form satellite
galaxies as bright as the LMC in Milky Way-sized haloes. Although in
the `vw4$\sigma$' model for the `Aq-D' halo there is a nearby galaxy
as bright as M33, it does not qualify as a satellite galaxy according
to our strict definition which requires satellites to be within a
radius of 280 kpc from the central galaxy. 

Away from the bright-end, the `vw' models (in which the wind
mass-loading increases towards smaller subhaloes) do reproduce the
observed satellite luminosity function reasonably well.  By contrast,
both of the `cw' models have overly steep luminosity functions with
too many faint satellites. The constant mass-loading wind model is not
able to suppress star formation in these faint satellites
sufficiently.  {We expect that if we had adopted the momentum-driven
wind model favoured by \citet{od06, od08}, in which the wind
mass-loading, $\eta_{\rm w}$, is proportional to $\sigma^{-1}$, we
would have obtained satellite luminosity functions whose faint-end
slopes are steeper than those in our `vw' models ($\eta_{\rm w}
\propto \sigma^{-2}$). On the other hand, it seems likely that if we had
not assumed that the winds are decoupled for a short period time, the
faint end of the resulting luminosity functions in the `cw' cases would
have been shallower than in our decoupled `cw' models \citep{ds08,owls}.}
Within any given halo, the `vw' models always match the satellite 
luminosity function better than the `cw' models.
From this point of view there is little to choose between the
`vw5$\sigma$' and `vw4$\sigma$' versions.

{The robustness of our results depends on how well the low-mass
end of the subhalo mass function is resolved in the simulations. In
order to identify the smallest resolved subhalo, we first compute the
cumulative subhalo abundance as a function of maximum subhalo circular
velocity, $N(> V_{\rm max})$. We then fit this cumulative velocity
function with a power-law, $N(> V_{\rm max}) \propto V_{\rm
max}^\alpha$; subhalo velocity functions in N-body simulations are
well represented by a power-law with $\alpha \sim -3$
\cite[e.g.][]{Diemand07b, aquarius}}

{
We define the smallest resolved circular velocity as the $V_{\rm max}$
value at which the velocity function first deviates by 10\% from the
power-law fit. We have computed this for the `vw5$\sigma$' and `cw700'
models of the Aq-D halo, which is the halo we use for convergence
tests later.  For the `vw5$\sigma$' model we find a value of
$12.9~{\rm km}~{\rm s}^{-1}$, which is consistent with the $V_{\rm
max}$ at which the subhalo velocity function begins to deviate from
that of the high-resolution counterpart (Aq-D-HR). The smallest
$V_{\rm max}$ of the subhaloes which contain at least 10 star
particles (i.e. the satellites) is 22.8~km~s$^{-1}$. We conclude that
the subhaloes that host satellites in the `vw' models are well
resolved.
For the `cw700' model, the minimum resolved circular velocity is
11.2~km~s$^{-1}$, which is consistent with the value for the
vw$5\sigma$ model. However, now the smallest $V_{\rm max}$ of the
satellites is 13.2~km~s$^{-1}$, which is close to the resolution
limit. Thus, in the `cw' models, the flattening seen at the very faint
end of the luminosity functions could be affected by numerical
limitations.}

\begin{figure}
\begin{center}
\includegraphics[width=8.5cm]{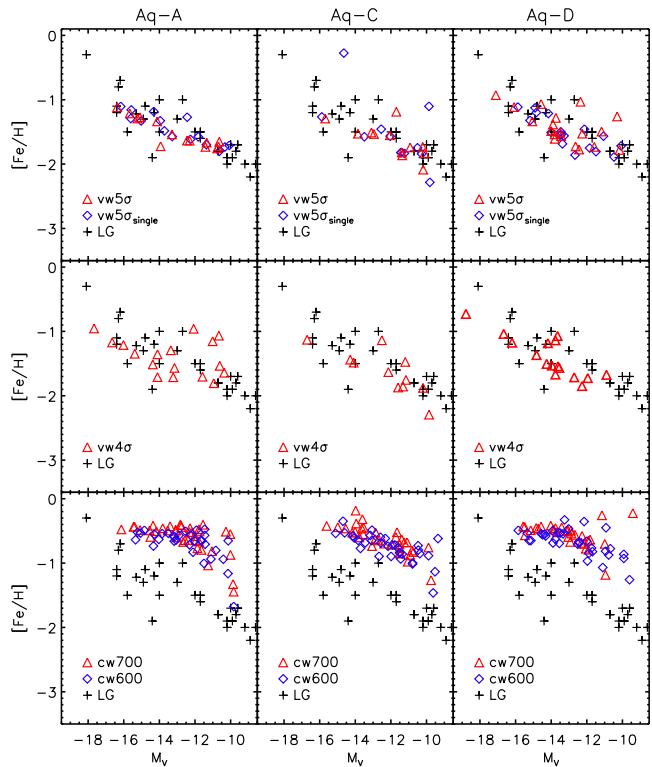}
\end{center}
\caption{The iron abundance relative to solar for 
  satellite galaxies, as a function of the absolute $V$-band
  magnitude. The panels are the same as in Fig.~\ref{morphology}. The
  abundances for the Local Group satellites are taken from
  \citet{mat98}, with the exception of the Magellanic Clouds which was
  taken from \citet{has95}.  Different symbols correspond to different
  wind models as indicated in the legend. The Local Group data are
  shown by plus signs.}
\label{lz}
\end{figure}

\subsubsection{Metallicity and abundance patterns}

We now investigate the metallicity of satellites which, in principle,
is also a sensitive diagnostic of the feedback physics
\citep[e.g.][]{ay87, fd08}.  In Fig.~\ref{lz}, we plot the iron
abundance as a function of satellite luminosity. (To compare our
simulations with the data, we adopt the values for the solar
abundances given by \citet{grev98}\footnote{Assuming the more recent
values for the solar abundances given by \citet{ags05} increases
[Fe/H] in the simulations only by $\sim 0.16$.}.) All of the `vw'
models reproduce the observed luminosity-metallicity relation for the
Local Group satellites quite well.  On the other hand, the iron
abundance of the `cw' satellites is almost a constant function of the
satellite luminosity, reflecting the constant wind mass-loading
factor.  These results strongly support models in which the wind
mass-loading scales as $\eta_{\rm w} \propto
\sigma^{-2}$. A momentum-driven wind would most likely produce a
relation intermediate between those of the `vw' and `cw' models, thus too
shallow to match the data, {while a coupled wind model would like
produce a steeper relation than the corresponding `cw' and `vw'
models. 
}

\begin{figure}
\begin{center}
\includegraphics[width=8.5cm]{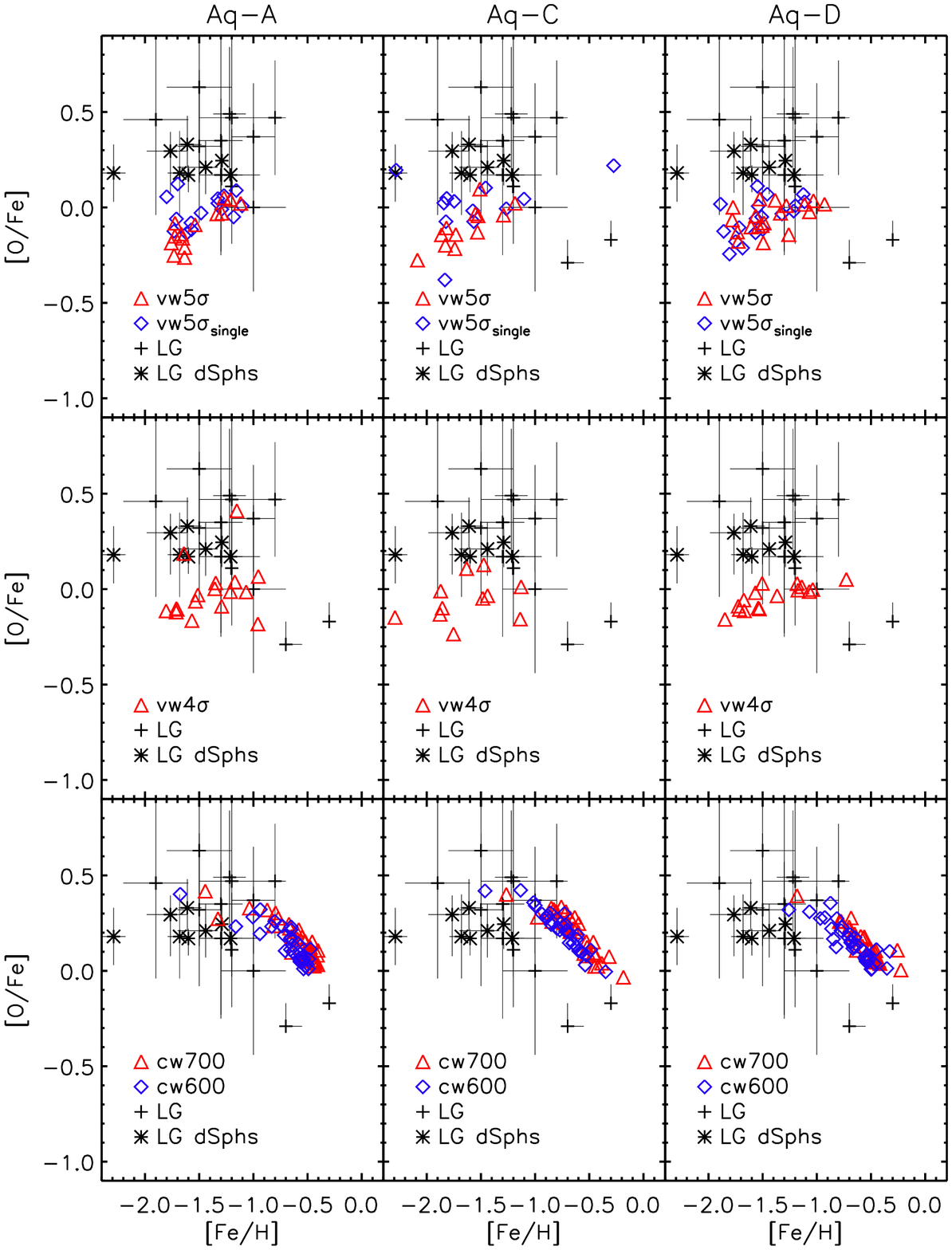}
\end{center}
\caption{The oxygen-to-iron abundance ratio of satellite galaxies 
  as a function of the iron abundance. 
  The panels correspond with those in
  Fig.~\ref{lz}. In addition to data from the sources given in
  Fig.~\ref{lz}, we plot the median values for Local Group dwarf
  spheroidals taken from \citet*{scs01} and \citet{she03}. For Mateo's
  data, the error bars represent $1 \sigma$ error, while for the
  Magellanic Clouds and dwarf spheroidals they show the larger of the
  25-75 percentile interval or the observational $1 \sigma$ error.}
\label{alpha}
\end{figure}

The alpha element-to-iron abundance ratio is often used as a measure
of the star formation timescale \citep[e.g.][]{no06}. Since SNe~Ia
hardly produce alpha elements, [O/Fe] decreases with time if a star
formation episode lasts longer than $\sim 1$Gyr, which is a typical 
SNe~Ia timescale. 
For an SSP of solar metallicity, it takes $\sim 0.9$ Gyr for SNe~Ia 
to produce as much iron as that produced by SNe~II\footnote{Assuming 
solar metallicity, the first SN~Ia goes off at $t \sim 75$ Myr in our 
model.}. In Fig.~\ref{alpha}, we show the oxygen-to-iron abundance 
ratio of satellite galaxies\footnote{Again, using more recent solar 
abundance determinations decreases [O/Fe] in the simulations by 0.17. 
However, the data in \citet{mat98} assume $12 +
\log($O/H$)_\odot = 8.93$, which is consistent with the values given by
\citet{grev98}} as a function of their iron abundance. Most of the
satellites in the `vw' models have [O/Fe] $\sim 0$, which is slightly
lower than the observed values. This might imply that the star
formation timescales in the simulated dwarf satellites are too long,
or that the model employed for the SN~Ia is incorrect
(e.g. \citealt{kob98}; our SN~Ia rate was calculated using the
scheme described by \citealt{gr98}, with parameters updated according to
\citealt{pcb98}), or that the assumed IMF is inaccurate
\citep[e.g.][]{naga05}. 
{Both the SN~Ia rate and the nucleosynthesis yields are highly
uncertain \citep[see e.g.][]{wie09b}. 
}
Exploring these issues is, however, beyond the scope of this
paper and we defer this to future work. 

The `cw' models reproduce the observed abundance patterns in the Milky
Way satellites reasonably well, even though they fail to match their
iron abundance (see Fig.~\ref{lz}). The reasons behind the different
patterns in the `vw' and `cw' are instructive. In the `vw' models, the
wind mass-loading is larger for smaller galaxies resulting in more
metal poor dwarfs (see Fig.~\ref{lz}); this, in turn, suppresses star
formation. However, the wind speed, $v_{\rm w} \sim 4$--$5 \sigma$, 
allows expelled gas to fall back later \citep[see][]{od08}, thus lengthening 
the effective star formation timescale.  
In the `cw' models, there is a critical velocity at which the wind speed 
is equal to the escape velocity from the halo, $v_{\rm esc}$.  For $v_{\rm w} 
\gg v_{\rm esc}$, the gas blown out of a halo never comes back and therefore 
the winds shorten the effective star
formation timescale. As a result, galaxies with small $V_{\rm max}$ 
become alpha-enhanced.  When the wind speed is of the order of
the escape velocity, the expelled gas is recycled as a fountain. Hence,
when $v_{\rm w} \lesssim v_{\rm esc}$, the effective star formation
timescale is longer than when $v_{\rm w} \gg v_{\rm esc}$ and the
oxygen-to-iron ratio begins to decrease relative to the value
set purely by SNe~II.

\begin{figure*}
\begin{center}
\includegraphics[width=16cm]{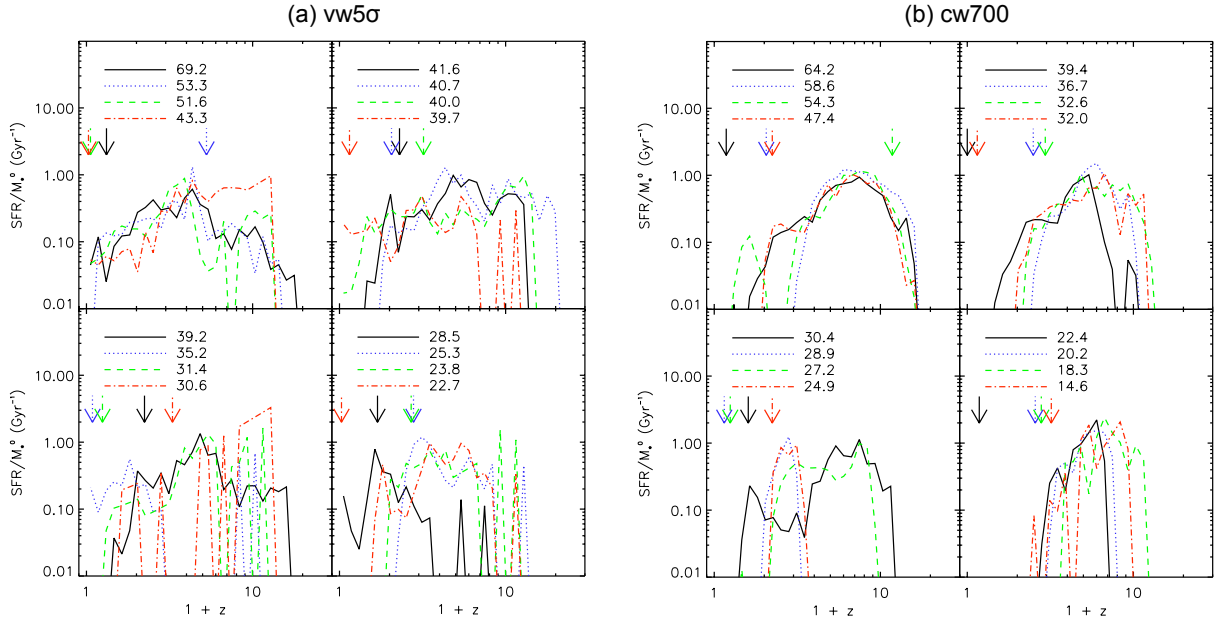}
\end{center}
\caption{(a): The star formation rate divided by $M_*^0$, the stellar
  mass at $z = 0$, for a selection of satellite galaxies in the
  vw5$\sigma$' model of the Aq-D halo. The labels give the value of
  $V_{\rm max}$ of each of the individual satellites, 
  in km s$^{-1}$.  The arrows indicate the redshift when the
  progenitors are accreted into a larger halo (i.e. when they become
  satellites). Note that we assume reionization occurs at $1 + z_{\rm
  re} = 10$.  (b): As (a) but for the `cw700' model of the Aq-D halo.}
\label{sfr_comp}
\end{figure*}

\subsubsection{Star formation histories}

The star formation histories of randomly selected satellites in two
models, `vw5$\sigma$' and `cw700' in the Aq-D halo, are shown in
panels (a) and (b) of Fig.~\ref{sfr_comp}, respectively. (We have
checked that the other models of each type are similar to these.) 
These plots allow us to investigate how reionization and infall into a
larger halo affect star formation. The arrows indicate the
`accretion' redshift, i.e. the moment when the main progenitor of a
satellite actually becomes a satellite galaxy by falling into a larger
halo.  The labels give the maximum of the circular velocity of the
selected satellites, $V_{\rm max}$, in km~s$^{-1}$, at $z = 0$.

We find that reionization (which happens at $z_{\rm re} = 9$) has little 
impact on the star formation histories of today's satellites. Even the 
smallest of them continue to make stars past this epoch. 
As shown by \citet{of09}, however, this is not the case for haloes that 
are too small at the epoch of reionization to host satellites. 
Haloes which have $V_{\rm max} \lesssim 12 $km~s$^{-1}$ at the epoch of 
reionization are unable to hold on to their gas which is evaporated by the 
UV photons and thus end up as dark subhaloes today. 

The figure shows that infall into a larger halo is important although
it does not always immediately truncate star formation (see also
\citealt{fon08}).
\cite{of09} demonstrate that infall tends to strip gas off the
infalling small object, eventually choking its star formation.  A
decrease in star formation soon after infall is apparent in some of
the examples in the figure, such as the haloes with $v_{\rm c}^{\rm
max} = 41.6$, 39.2, 28.5, 23.8${\rm ~km~s}^{-1}$ shown in panel (a)
for `vw5$\sigma$'. In a few cases, e.g. $v_{\rm c}^{\rm max} = 40.7$,
31.4, 25.3${\rm ~km~s}^{-1}$, star formation does cease completely
upon infall.

Similar remarks can be made regarding the satellites in the `cw700'
model. In this case, because the wind mass-loading is smaller than in
the `vw' models, subhaloes of smaller circular velocity can host
satellites. The truncation of star formation at the time of accretion
is more evident in satellites with such small circular velocities.  As
may be seen by comparing panel~(a) and~(b) of Fig.~\ref{sfr_comp}, at
a given $V_{\rm max}$, the star formation timescale is shorter
in the `cw700' model than in the `vw5$\sigma$' model. In addition, the
star formation in the `vw5$\sigma$' model is more variable in time
than in the `cw700' model. These differences arise from the fact that
a wind speed of $v_{\rm w} = 4-5 \sigma$ in the `vw' models can cause
wind particles to escape from a halo which, however, can be
re-accreted later on. By contrast, in the constant wind speed model,
re-accretion is inhibited whenever $v_{\rm esc} \ll v_{\rm w}$ while a
wind is suppressed whenever $v_{\rm esc} > v_{\rm w}$. As a result, the
star formation timescale is shorter than in the `vw' models 
whether $v_{\rm esc} \ll v_{\rm w}$ or $v_{\rm esc} > v_{\rm w}$.  
These differences in star formation
histories explain the differences in abundance patterns seen between
the two types of models (Fig.~\ref{alpha}).

\subsection{A high resolution simulation} 

\begin{figure}
\begin{center}
\includegraphics[width=8.0cm]{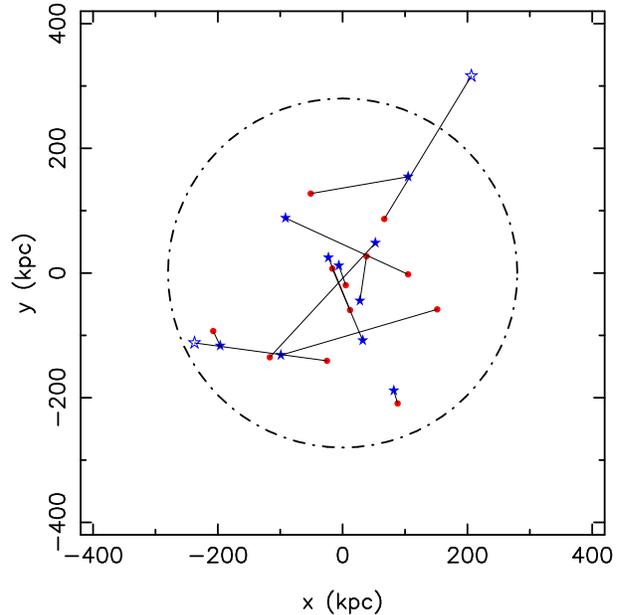}
\end{center}
\caption{ 
  The projected positions of the matched satellites in the two
  simulations with different resolution, Aq-D-HR and Aq-D.  The
  matching was carried out for satellites lying within 280 kpc from
  the centre that are brighter than $M_V = -11.2$, corresponding 
  to the luminosity of the faintest satellite within 280 kpc in the
  lower resolution simulation. Matched pairs are joined by thin
  lines. Circles denote satellites in Aq-D-HR and stars in Aq-D; open
  symbols indicate satellites found outside the 280 kpc sphere which
  is marked by the dot-dashed line.}
\label{matching}
\end{figure}  

In this subsection, we select the most promising wind model and
investigate its sensitivity to numerical resolution by resimulating it
with ten times higher mass resolution.  We selected the `vw5$\sigma$'
model on the grounds that it provides the best overall match to the
observed luminosity function and iron abundances of Local Group
satellites (although the oxygen-to-iron ratios, which are sensitive to
the adopted SN~Ia rate, are systematically low).  The run, Aq-D-HR, is
detailed in Table~\ref{resolution} (thanks to its low concentration,
the Aq-D halo is the least computationally expensive of the three).
Since we only consider the `vw5$\sigma$' model in this subsection, for
simplicity, we refer to the high and low resolution simulations as
Aq-D-HR and Aq-D (or simply as the high and low resolution
simulations), respectively.

{We first examine the correspondence between the positions of
individual satellites in the two simulations. To do this, we trace the
dark matter particles of a particular subhalo in one of the
simulations back to the initial conditions and search for a matched
subhalo in the other simulation \citep[see][]{aquarius}. We include
all the satellites lying within 280~kpc in the low resolution
simulation and the satellites in the high resolution simulation which
also lie within 280~kpc and are brighter than $M_V = -11.2$,
corresponding to the faintest luminosity of satellites within 280~kpc
in the low resolution simulation.}

\begin{figure}
\begin{center}
\includegraphics[width=8.5cm]{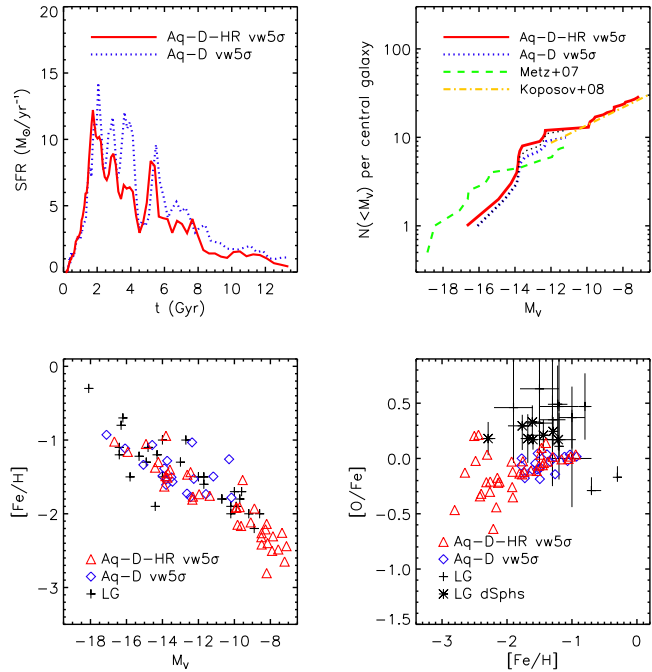}
\end{center}
\caption{Comparison of the low and high resolution simulations of the
`vw5$\sigma$ model in the Aq-D halo. In all panels, the high
resolution (Aq-D-HR) case is shown by solid lines and the low
resolution (Aq-D) case by dotted lines. {\it Upper-left:} star
formation histories of the central galaxies, as in
Fig.~\ref{sfh}. {\it Upper-right:} the satellite luminosity functions
within 280~kpc, as in Fig.~\ref{lf}. The thin black dotted line shows
the luminosity function of the low-resolution satellites that are
matched to the bright ($M_V \ge -11.2$) high-resolution satellites.
{\it Lower-left:} the iron abundance as a function of satellite
luminosity, as in Fig.~\ref{lz}. {\it Lower-right:} oxygen-to-iron
abundance as a function of the satellite luminosity, as in
Fig.~\ref{alpha}.}
\label{highres}
\end{figure}

{
The result is shown in Fig.~\ref{matching}, where the matched pairs
are linked by a line.  The mean spatial offset between matched pairs
is $\sim 100$ kpc, somewhat larger than the mean offset of $\sim 54$
kpc found between matched pairs in Aq-A-4 and Aq-A-1\footnote{Aq-A-4
has similar resolution to Aq-D-HR and Aq-A-1 has $\sim 100$ times
better resolution than Aq-D-HR.} \citep{aquarius}.  This difference is
not due to small number statistics - we have only 12 pairs of
satellites - since a similar offset is found for matched pairs of all
subhaloes within 280~kpc, whether or not they host satellites. The
mismatch is more likely due to slight orbital phase deviations - which
translate into large spatial separations - produced by differences in
the potential of the two galaxies whose evolution is not identical, as
we will show in Fig.~\ref{highres}. Slight difference in the mass of the
subhaloes due to baryonic processes may also contribute to this
offset. In spite of the relatively large spatial offsets, the
photometric and chemical properties of the satellite populations in
the two simulations are remarkably similar, as we will show later in this
subsection.
}

{In Fig.~\ref{highres}, we present the results of the convergence
tests. We first compare the star formation histories in the central galaxies 
in two simulations in the upper-left panel. Although the shape of the two
curves is very similar, the higher resolution simulation has a
slightly lower mean star formation rate. This likely reflects
enhanced mass loss through winds from small haloes not resolved in the
lower resolution simulation. At $z=0$, the stellar mass in the central
galaxy is 18\% lower in the Aq-D-HR simulation than in the Aq-D
simulation.  }

The upper-right panel shows the satellite luminosity functions
obtained in the two simulations. They agree very well. The slight
differences between them arise mainly from the requirement that the
samples should contain only satellites found within 280~kpc of the
central galaxy. Two of the satellites in the high resolution
simulation have counterparts in the low resolution simulation which
lie outside the 280~kpc radius. One of these can be clearly seen in
Fig.~\ref{matching}; the other one is projected inside the circle.
When we include these two satellites, the luminosity functions in the
low and high resolution simulations agree even better, down to the
faintest satellite luminosity in the Aq-D ($M_V \simeq -11$). 
This suggests that our definition of satellites is robust. 
Encouragingly, the faint end of the satellite luminosity function 
in the high resolution simulation matches that of the newly discovered 
SDSS satellites \citep{kop08} remarkably well. This provides further 
support for the `vw' winds in which 
$v_{\rm w} \propto \sigma$ and $\eta_{\rm w} \propto \sigma^{-2}$.

{We again estimate the minimum resolved circular velocity of the
subhaloes in the high-resolution simulation as described in Section
\ref{section-lf}. The minimum resolved value is 6.1~km~s$^{-1}$, while
the minimum value of the subhaloes that host a satellite is
12.1~km~s$^{-1}$. Hence, all the subhaloes hosting satellites are well
resolved in the simulation.  The minimum $V_{\rm max}$ of the
satellites in these models is set by reionization \citep{ogt08,
of09}. }

In the lower panels of Fig.~\ref{highres}, we compare the chemical
properties of the satellites in the two simulations. Both the iron
abundance and the oxygen-to-iron ratios are consistent at the two
resolutions.  The high resolution simulation predicts that the
luminosity-metallicity relation for satellites should extend down to
at least $M_{\rm v} \sim -7$ (lower-left panel).  There is a hint in
the lower-right panel that the oxygen-to-iron abundance ratio in the
simulated satellites increases slightly with metallicity. This trend
is possibly inconsistent with the Local Group dwarf spheroidal data,
in which the alpha-to-iron abundance ratios appear to be almost
constant or slightly increasing with decreasing metallicity
\citep[e.g.][]{tht09}.

Comparing matched pairs individually, we find that the stellar mass in
the high-resolution satellites is, on average, 30\% higher than in
their low-resolution counterparts. (This trend is opposite to that for
the central galaxy where the high-resolution simulation forms a less
massive galaxy.) The metallicity in the high-resolution satellites is,
on average, only 8\% higher than in the low-resolution simulation. 

\begin{figure}
\begin{center}
\includegraphics[width=8.5cm]{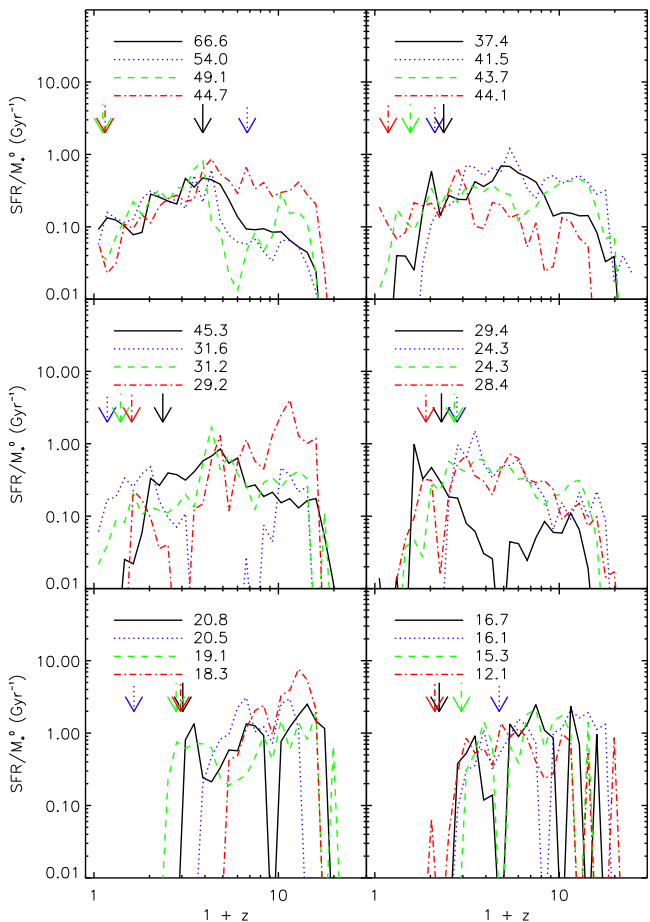}
\end{center}
\caption{As (a) of Fig.~\ref{sfr_comp}, but for the `vw5$\sigma$' model in the
Aq-D-HR halo. The top and middle panels show star formation histories 
of matched high-resolution counterparts of the low-resolution satellites 
shown in (a) of Fig.~\ref{sfr_comp}. The bottom panels show star formation histories 
of satellites that are not resolved in Aq-D.}
\label{sfrhres}
\end{figure}

Finally, in Fig.~\ref{sfrhres}, we show the star formation histories
of a selection of satellites in the high-resolution simulation.  The
top and middle panels show the star formation histories of the matched
counterparts shown in panels (a) of Fig.~\ref{sfr_comp}. These
histories are qualitatively similar to those seen in the low
resolution simulation (Fig.~\ref{sfr_comp}), indicating that the time
variability is not a resolution effect but a result of the winds.  The
sporadic nature of the star formation makes the effective star
formation timescale longer (and lowers the alpha-to-iron ratio) in the
`vw' models compared to the `cw' models.  
{As we saw earlier, the
star formation histories of the surviving satellites retain no obvious
imprint of reionization. Even the satellites in the bottom panels of
Fig.~\ref{sfrhres}, which were not resolved in the low-resolution
counterpart (Aq-D), continue forming stars past the reionization
epoch. We note, however, that the gas content in subhaloes that has 
failed to form stars was evaporated at the epoch of reionization 
\citep{of09}. This process is likely to be responsible for the suppression 
of cosmic star formation rate at the epoch of reionization found by 
larger cosmological simulations \citep{gimic, owls}.} 
It is interesting to note that star formation in these
smallest satellites ($V_{\rm max} \lesssim 20$ km~s$^{-1}$) ceases at
the time of accretion or earlier.  We discuss in detail these sorts of
effects on surviving satellites and failed satellites (i.e. subhaloes
that do not host stars) in a companion paper
\citep{of09}.  

In summary, our high resolution resimulation of the `vw5$\sigma$'
model shows that the statistical properties of the satellite
population in our low resolution calculations are robust to a change
in mass resolution of a factor of ten. Furthermore, the high
resolution simulation allows us to extend the model luminosity
function by 4 magnitudes, thus enabling a comparison with recent
data.

\section{Summary and discussion} 

We have performed cosmological hydro-chemodynamic simulations of
galaxy formation in 3 Milky Way-sized haloes taken from the Aquarius
project \citep{aquarius}. A primary aim of our study has been to
constrain the nature of some of the feedback processes that must have
operated during galaxy formation. We have done this by focusing on
satellite galaxies whose shallow potential wells make them
particularly sensitive to feedback effects. As a byproduct, we have been
able to explore some of the controversial properties of satellite
galaxies, such as their luminosity function, in the context of the
$\Lambda$CDM cosmology which we have assumed in our simulations.

Although our simulations are amongst the largest of their kind
performed to date, they lack the resolution to follow the physics of
the interstellar medium {directly}.  
{Such processes must be
included as `sub-grid physics'. In particular, we have assumed that
energy injected by SNe generate galactic winds. We have assumed
further that all the energy released by SNe is deposited as kinetic
energy in the winds.}

{
Early studies, as well as more recent ones have shown that many
observed galaxy properties cannot be reproduced unless most of the SN
energy is used to blow gas out of galaxies \citep[e.g.][]{sh03b,
oka05, od08, gimic, owls}. Since a large fraction of SN energy must be
radiated away, other sources of energy seem to be required to power
strong winds such as radiation from young stars
\citep{mqt05, ns09}. In this study we have included these unresolved
processes in a purely phenomenological way by assuming two types of
`energy-conserving' winds. }
In the first type, the initial wind speed is taken to be proportional
to the local velocity dispersion of the dark matter, $v_{\rm w}
\propto \sigma$, {as suggested by recent data \citep{mar05}}, and
thus, the wind mass-loading, $\eta_{\rm w} \propto \sigma^{-2}$ (`vw'
models). In the second type, the initial wind speed and the wind
mass-loading are assumed to be constant for all galaxies (`cw'
models), {as suggested by earlier data
\citep{mar99}}. Disc-dominated galaxies formed in several of our 
simulations.

In some of our simulations, we included a multiphase model for
star-forming gas, but we found that this made little difference to the
outcome \citep[see also][]{owls}, apart from the overall morphology of the 
central galaxy. The use of a stiff equation of state for the ISM tends to 
promote the formation of disc-dominated galaxies by stabilising gaseous discs
against gravitational instability. The key to the diversity of behaviours that 
we find is not the treatment of the ISM, but rather the treatment of the 
SN-driven winds.

The `vw' models give a reasonable match to the observed luminosity
function of Local Group satellites. A major factor in this success is
the behaviour of the mass-loading in the wind which becomes
increasingly large in smaller galaxies. If the mass-loading is kept
constant as in the `cw' case, galaxy formation is not sufficiently
suppressed in small haloes resulting in a satellite luminosity function
which rises much too steeply at the faint end. Previous SPH
simulations had already shown that an acceptable match to the
abundance of bright satellites can be obtained from $\Lambda$CDM
initial conditions
\citep{lib07, gov07, mac09}, thus confirming the conclusion from
early semi-analytic studies that the so-called `satellite problem' in
the CDM cosmology \citep{kly99, moo99} disappears when proper
account is taken of the baryonic processes involved in galaxy
formation \citep{bkw00, ben02, som02}. Our high-resolution simulation, 
however, extends the comparison with the luminosity function to much 
fainter magnitudes than previous SPH simulations.

The `vw' feedback model that produces an acceptable satellite
luminosity function, also produces an acceptable
luminosity-metallicity relation for the satellites. By contrast, the
`cw' model fails to reproduce this relation for the same reasons that
it fails to reproduce the luminosity function: the constant
mass-loading leads to an iron abundance in satellites with $v_{\rm
esc} \gg v_{\rm w}$ which is essentially independent of luminosity,
contrary to what is observed.

None of our `vw' models reproduces the oxygen-to-iron abundance ratio
measured for dwarf satellites.  There are several possible reasons for
this discrepancy: the star formation timescale may be too long, the
model of SNe~Ia or the IMF that we have adopted may be incorrect, or
our assumed oxygen yield may be too low. Given the large uncertainties
in the value of the yield, the IMF and SNe~Ia rates, we do not
consider this shortcoming of our models to be fatal although further
work is required. For example, a top-heavy IMF in these metal-poor
objects could provide the solution although it is unclear whether such a
radical assumption would destroy the agreement of the model with other
observables.

{Based on our estimate of the minimum resolved circular velocity
of subhaloes in the simulations, we find that subhaloes hosting
satellites in the `vw' models are reliably resolved down to the
faintest magnitude. Thus, the faint-end slope of the satellite
luminosity function and the luminosity-metallicity relation in the
`vw' models are robust. This conclusion is confirmed by a direct
convergence study of one model (`vw5$\sigma$' of the Aq-D): low- and
high-resolution simulations give consistent results. On the other
hand, in the `cw' models in which the smaller wind mass-loading factor
allows satellites to form in halos with small values of $V_{\rm max}$,
the faint-end slope of the satellite luminosity function could be
affected by limited numerical resolution although we estimate that
any such effects would be restricted to the extreme faint end. }

Barring the alpha-to-iron ratio, we conclude that an energy-conserving
wind model in which the mass loading scales inversely with square of 
velocity dispersion provides a viable model for feedback as judged by the
properties of satellite galaxies. Our preferred wind model is
different from the `momentum-driven' wind models favoured by 
\cite{fd08} in order to explain the mass-metallicity relation of
large galaxies. In this kind of model, the wind speed is proportional
to $\sigma$ as in our `vw' models, but the mass-loading scales as
$\eta_{\rm w} \propto \sigma^{-1}$ (rather than as $\eta_{\rm w}
\propto \sigma^{-2}$). It could be that the mechanisms that drive
galactic outflows in small and large galaxies are different.

In `vw' models with wind speed, $v_{\rm w} \simeq 4-5\sigma$, the mass
loading is large in small galaxies. Winds therefore remove substantial
amounts of star-forming gas, but much of this material eventually
falls back onto the galaxy. As a result, the star formation in small
satellites is episodic and has a much longer timescale than in the
`cw' models in which the gas is expelled from the small halo, never to
return. These general properties underlie the different predictions
for the number density of faint galaxies and their chemical properties
in the different models.

The star formation histories of the satellites retain no obvious
imprint of the reionization of gas at early times. Accretion onto
larger haloes, on the other hand, often affects subsequent star
formation and, in low mass satellites ($V_{\rm max} \lesssim 20~{\rm
km}~{\rm s}^{-1}$), it truncates it altogether. 
{Only 5\% of
subhaloes whose circular velocities are higher than the minimum
resolved value ($V_{\rm max} > 6.1~{\rm km}~{s}^{-1}$) in the
high-resolution simulation (Aq-D-HR) host satellites.  The vast
majority of subhaloes do not manage to make a visible galaxy and
remain dark.} The mechanisms that distinguish visible from dark
satellites are investigated in a companion paper \citep{of09}.

\section*{Acknowledgements}

We are grateful to Volker Springel for providing us with the {\small
GADGET-3} code and to Joop Schaye, Claudio Dalla Vecchia, and Rob
Wiersma for providing us with tabulated radiative cooling and heating
rates. Our simulations were performed at the Center for Computational
Sciences in the University of Tsukuba, the Cosmology Machine at the ICC,
Durham, and the Cray XT4 at CfCA of NAOJ. This work was supported in
part by the {\it FIRST} project based on Grants-in-Aid for Specially
Promoted Research by MEXT (16002003), Grant-in-Aid for Scientific
Research (S) by JSPS (20224002), and an STFC rolling grant to the
ICC. 
TO acknowledges financial support by Grant-in-Aid for Young Scientists 
(start-up: 21840015). 
CSF acknowledges a Royal Society Wolfson research merit award.

\bsp
\label{lastpage}
\end{document}